\documentclass[article,12pt,showpacs,preprintnumbers,amsmath,amssymb]{revtex4}%
\usepackage{color}
\usepackage{graphicx}

\begin{document}

\preprint{}

\title{Random $3D$ Spin System Under   the
 External Field and Dielectric Permittivity Superlattice Formation }

\author{Ashot S. Gevorkyan$^{1,2,3}$}

\author{Chin-Kun Hu$^{1}$}

\affiliation{$^1$Institute of Physics, Academia Sinica, Nankang,
Taipei 11529, Taiwan}\affiliation{$^2$Institute of Applied
Problems of Physics, NAS of Armenia, Nerssian St. 25, Yerevan
375014, Armenia}

\date{\today}

\begin{abstract}
A dielectric medium consisting of roughly polarized molecules is
treated as a $3D$ disordered spin system (spin glass). A
microscopic approach  for the study of statistical properties of
this system on micrometer space scale and nanosecond time scale of
standing electromagnetic wave is developed.  Using  ergodic
hypothesis the initial $3D$ spin problem  is reduced to two
separate $1D$ problems along  external field propagation. The
first problem describes the disordered spin chain system while the
second one describes a disordered $N$-particle quantum system with
relaxation in the framework of Langevin-Schr\"odinger (L-Sch) type
equation. Statistical properties of both systems  are investigated
in detail. Basing on these constructions, the coefficient of
polarizability, related to collective orientational effects, is
calculated. Clausius-Mossotti formula for dielectric constant is
generalized.  For dielectric permittivity function generalized
equation is found taking into account Clausius-Mossotti
generalized formula.
\end{abstract}
 \maketitle

\section{Introduction}

The formation and control of periodic nanostructures in various
type media (Media with Periodically Modulated Refractive Index
(MPMRI)) are one of most important applied problems. First of all
it is related to the possibility of the creation of compact UV or
X-ray Free-Electron Lasers (FEL) based on the principle of
transition radiation (TR) (see for example \cite{Apol}). Currently
the following two problems are discussed intensively:
\begin{enumerate}
\item A gas-plasma medium with periodically varied ionization
density
\cite{chen,Agrawal,Nakajima,Bazelev,Bazelev1,Karbushev,Feodorov,Artem,Zhang},
\item A specially periodical solid-state superlattice-like (SSL)
structures, which are composed of layers with different refracted
indices
\cite{Piestrup,Reid,Bekefi,Kaplan,Dubovikov,Liu,Piencus,Piestrup1,Kapl1}.
\end{enumerate}
TR is generated due to the difference in frequency-dependent
dielectric constants (dielectric permittivity functions)  of
adjacent layers (recall that radiation power is proportional to
$[\epsilon_1(w)-\epsilon_2(w)]^2$). Therefore, an important
problem is the possibility of the control  of this difference by
means of some external field. In other words, a problem is to
construct a superlattice with dielectric constant difference
between neighboring domains having a form
$[\epsilon_1(w,{\bf{g}})-\epsilon_2(w,{\bf{g}})]^2$, where
${\bf{g}}$ describes controlling parameters,
$\epsilon_1(w,{\bf{g}})$ and $\epsilon_2(w,{\bf{g}})$ are
dielectric permittivity functions in the neighboring regions.
According to theoretical and experimental studies the periodical
structures can be created in condensed matter by means of external
electromagnetic or acoustic fields (see in particular
\cite{Morozov,Rasporin,Mingrong,Xue-Zhang}).

This idea have been applied recently for experimental TR
generation \cite{Al}. In particular, it was shown that the
electron beam of 20$Mev$, passing through amorphous silicon
dioxide $a-SiO_2$ with the standing electromagnetic wave (of 10
$GHz$ frequency and 3 $\mu m$ wavelength) inside, produces anomaly
high radiation. Preliminary investigations explain its high
intensity as a result of the multiple passing of the electron beam
through  interfaces between regions  with different dielectric
constants. The theoretical study explained the appearance of
one-dimensional superlattice order in random media by media
polarization, caused by orientation relaxation of elastic
molecular dipoles in external field \cite{Ash}.

The main goal of this paper is systematic investigation of the
relaxation processes and critical effects in the disordered $3D$
spin system  type of $a-SiO_2$ under the influence of external
electromagnetic field, which forms a standing wave in the medium.
In particular, we develop a mathematical approach for description
of statistical behavior of disordered $3D$ spin system along the
direction of  standing wave propagation on space-time scale of its
period.

 In the framework of this approach we generalize Clausius-Mossotti
 equation for dielectric constant  and the equation
for frequency-dependent dielectric permittivity on the scale of
space-time period of standing electromagnetic wave.

\section{Formulation of the problem. New ideas and basic formulas}

Let us begin by discussing the fundamental problem of  the
dielectric constant space-time modulation (generation of
dielectric constant superlattice with controlling parameters) in
some types of amorphus materials.

 In this subsection, we shall give mathematical formulations for
dielectric constant in the presence of an standing low electrical
field. A particular attention will be pointed to investigation of
dispersion properties of dielectric permittivity function where
generation of collective orientation effects is possible under the
standing low electromagnetic field.

The starting point in our discussion will be Clausius-Mossotti
relation for  dielectric constant. It is known that in isotropic
mediums (as well as in the crystals with cubic symmetry) the
dielectric constant is well described by the Clausius-Mossotti
formula (see for example   \cite{Kit,Grif,Becker})

\begin{equation}
\frac{\epsilon_{s}-1}{\epsilon_{s}+2}=\frac{4\pi}{3}\sum_{m}N^{0}_{m}
\alpha^{0}_{m}, \label{01}
\end{equation}
where $N^{0}_{m}$ is the concentration of particles (electrons,
atoms, ions, molecules (or dipoles)) with the given $m$ types of
polarizabilities and $\alpha^{0}_{m}$ correspondingly are
coefficients of polarizabilities.  From this formula it follows
that the static  dielectric constant $\epsilon_{s}$ depends on the
polarizabilities properties of the particles as well as on their
topological order. In the external alternating field the
homogeneity and the isotropy of the medium is often lost. Then, it
is expected that the formula (\ref{01}) will be applicable after
the minor generalization.

The object of our investigation will be solid state dielectrics
type of amorphous silicon dioxide $a-SiO_2$. According to
numerical $ab$ $initio$ simulations \cite{Tu}, the structure of
this type compound is well described by $3D$ random network (see
Fig. 1.)

\begin{figure}[h]
\begin{center}\includegraphics[height=60mm]{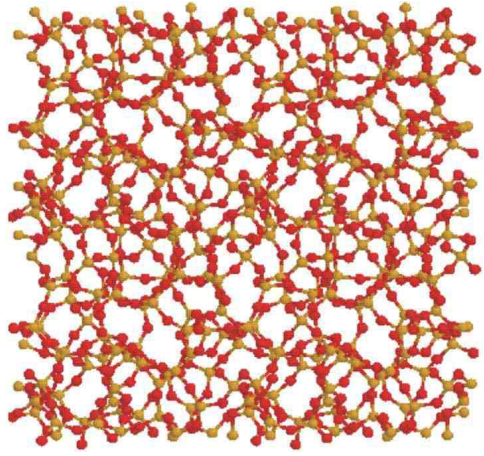}\end{center}
\caption{ \emph{The structure of amorphous silicon dioxide
($a-SiO_2$) is described by $3D$ random network with covalent
bonds. Every silicon vertex (gold sphere) has 4 edges and every
oxygen vertex (red sphere) has 2 edges.}}
\end{figure}

The white and black nodes in this figure correspond to different
atoms while links between them correspond to covalent bonds. The
redistribution of charges in outer electronic shells takes place
because of the asymmetry of the bounded atoms. As a result some
atoms acquire positive charge while others acquire negative ones.
Thus this type of  compound can be treated as a disordered $3D$
rigid dipoles system (see Fig. 2). Below this system will be
called a $3D$ disordered spin system. For the description of
amorphus media we use $3D$ lattice   with the lattice constant
$d_{0}(T)=\{m_0/\rho_{0}(T)\}^{1/3}$, where $m_0$ is the molecule
mass, $\rho_{0}$ is the density and $T$ is the temperature. The
lattice contains one random spin per elementary cell. Note that it
has random direction as well as random location inside the cell.
\begin{figure}[h]
\begin{center}\includegraphics[height=60mm]{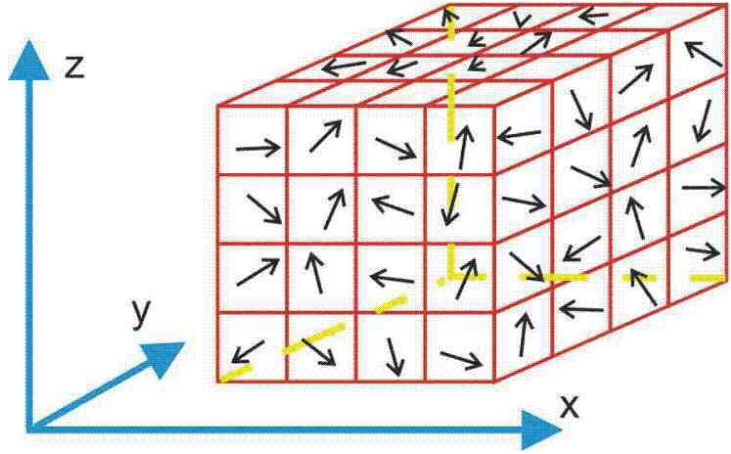}\end{center}
\caption{ \emph{The amorphous silicon dioxide (see Fig. 1.) maybe
represented as a $3D$  lattice where in every cells randomly put
one rigid dipole (spin).} }
\end{figure}
Suppose now that with the help of external electromagnetic filed a
standing wave is formed in the medium:

\begin{equation}
E(x;E_{0},\Omega,\varphi_{0})=2E_{0}\sin(\varphi_{0})cos(kx),
\quad \varphi_{0}=\Omega t_{0}, \label{02}
\end{equation}
where $\varphi_{0}$  and  $t_0$  are correspondingly the initial
phase and time,  $\Omega$ is the wave frequency,
$k=2\pi/\lambda_s$ and $\lambda_s$ is the wavelength (see Fig. 3).
\begin{figure}[h]
\begin{center}\includegraphics[height=60mm]{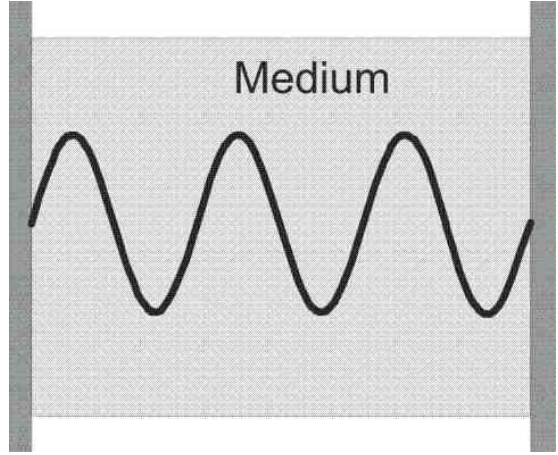}\end{center}
\caption{a) \emph{Disordered $3D$ medium under the standing
electromagnetic wave. } }
\end{figure}
 The following natural question arises. How does dielectric
constant change on scale of wavelength period and  on time
interval $\Delta{t}\ll{\Omega^{-1}}\sim{10^{-9}}sec$ when the
relaxation time of molecular dipoles is
$\tau\sim10^{-11}\div10^{-12}sec\ll{\Omega^{-1}}$. This question
is important because faster processes, like the transition or
Cherenkov radiation, can go in the media. Note, that the time of
relativistic electron is passing the  wavelength
($\lambda_s\sim{10^{-4}}cm$) of standing wave and the time of
formation of transition or Cherenkov photons in this layer is
smaller than  $10^{-15}sec $. This time interval is essentially
lower than the time during which the standing wave is stationary.
Since the wavelength is supposed to be much bigger than the
inter-dipole distance $\lambda_s\gg d_0$, the Clausius-Mossotti
relation still remains true. In this case the main problem is to
calculate the polarizabilitiy coefficient related to orientational
effects.

Taking into account the external field, one can express the
polarization of the matter at arbitrary point as the macroscopic
self-consistent relation:

\begin{equation}
\vec{P}(\vec{r},d_0(T))=\sum_{\vec{l}}\vec{p}\,(\vec{l}-\vec{r})=\sum_{\vec{l}}
\Bigl[\sum_m{n}_m{\alpha_m(\vec{l}-\vec{r})}\vec{E}_{loc}(\vec{l}-
\vec{r})\Bigr], \quad \vec{l}\equiv \vec{l(l_x,l_y,l_z)},
\label{03}
\end{equation}
where $\vec{l}$ is the vector of $3D$ lattice, $\vec{p}$
correspondingly the dipole moment of molecule. The second equation
in (\ref{03})  contributes in the value of the dipole moment
(spin). Note, that the number of the carriers of given
polarization type in elementary cell is $n_m\sim({d_0(T)})^{-3}$,
$\alpha_m$ coefficients of corresponding types polarizabilities
taking into account the external field and $\vec{E}_{loc}$ is
local field, the effective field, at the site of an individual
molecule that causes the induced polarization. Each effect adds
linearly to the net dipole moment per molecule which is a fact
verified by experiments. Under the action of external field
different polarization types arise in media. However as the simple
analysis shows the value of polarizabilities coefficient
determined by the orientational effects is essentially higher than
others.

Note that the elastic orientational polarizability coefficient in
amorphous media $\alpha_{dip}(\vec{l}-\vec{r})$  randomly depends
on the cell location. This fact is  the consequence of random
orientation of the local field strengths  $\vec
E_{loc}(\vec{l}-\vec{r})$  with respect to the external field
$\vec{E}(x;E_0,\Omega,\varphi_0)$. Therefore all the terms in the
right  side of  (\ref{01}) are basically known and well studied in
literature (see for example \cite{Kit,Grif,Becker}) except from
those which are  connected with orientational effects.

The orientational effects have a collective nature and are
characterized by average value of random sum
 $\sum_{\vec{l}}\,\alpha_{dip}(\vec{l}-\vec{r})$ (sum of random
coefficients of orientational polarizabilities).

Multiplying both sides of the relation (\ref{03}) on the external
field, we obtain:

\begin{equation}
\vec{P}(\vec{r},{\bf{g}})\vec{E}(x,{\bf{g}})=
-\delta{U(\vec{r},{\bf{g}})}= \sum_m{n}_m\Bigl[\sum_{\vec{l}}
{\alpha_m(\vec{l}-\vec{r})}\vec{E}_{loc}(\vec{l}-\vec{r})\Bigr]\vec{E}
(x;E_0,\Omega,\varphi_0), \label{04}
\end{equation}
where $-\delta{U(\vec{r},{\bf{g}})}$ describes the potential
energy of amorphous matter in the external field, symbol
${\bf{g}}$ shows   parameters of standing wave
$(E_0,\Omega,\varphi_0)$ (controlling parameters).  Later we will
consider the statistical properties of the medium in the direction
of wave propagation.

Taking into account  formula (\ref{04}), one can obtain the
following expression for the part of potential energy of the $3D$
spin system, which is related with orientation effects in external
field:
\begin{equation}
-\delta{U_{dip}(\vec{r},{\bf{g}})}=\sum_{\vec{l}}
{\alpha_{dip}(\vec{l}-\vec{r})}\vec{E}_{loc}(\vec{l}-\vec{r})
\vec{E}(x;E_0,\Omega,\varphi_0). \label{05}
\end{equation}

Let us pick out a layer with  volume $V=L_x\times L_y\times L_z$
in the infinite crystal lattice, where $L_x\sim(\lambda_s)\gg
d_0(T)$ and $(L_y,L_z)\rightarrow(\infty,\infty)$. It is easy to
see that this volume is filled with infinite number of $L_x$-site
random steric spin chains ( see Fig. 4).
\begin{figure}[h]
\begin{center}\includegraphics[width=0.7\textwidth]{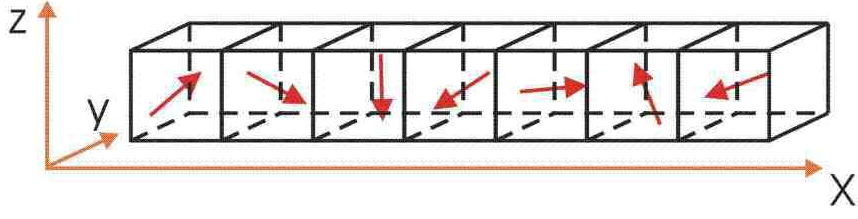}\end{center}
\caption{ \emph{The steric $1D$ random spin chain system}. }
\end{figure}

 An important problem is now to calculation the mean value of
the interaction potential between  the spin layer and external
field. Formally for it the following expression may be written:
$$
-\delta U_{V}(\vec{r},{\bf{g}})=
-\sum_{\vec{l}_{\perp}}\delta{U_{L_x}}(\vec{l}_{\perp}|
\vec{r},{\bf{g}}),
$$

\begin{equation}
-\delta{U_{L_x}}(\vec{l}_{\perp}|\vec{r},{\bf{g}})=
\sum_{l_x}{\alpha_{dip}(\vec{l}-\vec{r})}\vec{E}_{loc}(\vec{l}-\vec{r}
)\vec{E}(x;E_0,\Omega,\varphi_0), \qquad
\vec{l}_{\perp}\equiv\vec{l}_{\perp}(l_x,l_y),\label{06}
\end{equation}
where -$\delta{U_{L_x}}(\vec{l}_{\perp}|\vec{r},{\bf{g}})$ is
 the interaction potential between the steric spin chain
and external field. First, we take the mean value of the potential
$-\delta{U_{V}}(\vec{r},{\bf{g}})$ on $(y,z)$ plane:
$$
\lim_{S_{\perp}\longrightarrow\infty}\frac{1}{S_{\perp}}\int
\delta{U_{V}}(x,{\bf{g}})\,dS_{\perp}=
\lim_{S_{\perp}\longrightarrow\infty}\frac{1}{S_{\perp}}\int
\sum_{\vec{l}_{\perp}}\delta{U_{L_x}(\vec{l}_\perp|x,{\bf{g}})}
\,dS_{\perp}
=\langle{\delta{U_{L_{x}}}(x;{\bf{g}})}\rangle_{(\updownarrow)},
$$
where  $S_\perp={L_y}\times {L_z}$ and
$\langle...\rangle_{(\updownarrow)}$ is averaging over all
possible stable steric $1D$ spin configurations. The integral
corresponds to the potential energy density
-$\langle{\delta{U_{L_{x}}}(x;{\bf{g}})}\rangle_{(\updownarrow)}$
on $(y,z)$  plane depends on the distance $x$. Taking into account
the fact that the distribution of spins in the plan $(y,z)$ is
random but isotropic (see Fig. 5) it is simple  to prove that in
the limit of $S_\perp\to\infty$ it becomes the full self-averaging
of spin system.
\begin{figure}[h]
\begin{center}\includegraphics[height=60mm]{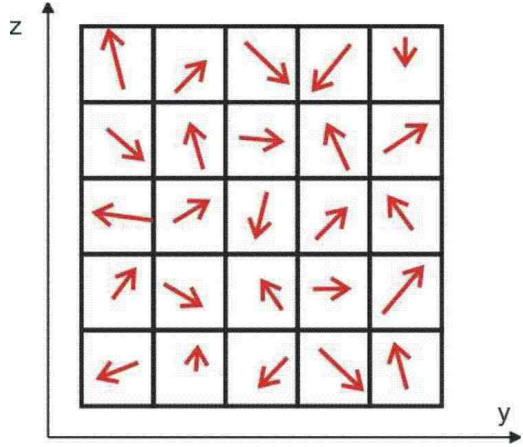}\end{center}
\caption{ \emph{The projection of $3D$ disordered spin system on
the plan $(y,z)$. }}
\end{figure}

 This implies that one may use  Birgoff ergodic
hypothesis \cite{LGP} and averages by plan $(y,z)$ interaction
potential
-$\langle{\delta{U_{L_{x}}}(x;{\bf{g}})}\rangle_{(\updownarrow)}$
  obtains also by the integration of chain's intrinsic
  energy distribution:

\begin{equation}
\langle{\delta{U_{L_{x}}}(x,{\bf{g}})}\rangle_{(\updownarrow)}=
\frac{\int_{-\infty}^0 {\delta{U_{L_{x}}}({\bf{E}}|x,{\bf{g}})}
{\bf{Z}}({\bf{E}};{\bf{g}}) d{\bf{E}}} {\int_{-\infty}^0
{\bf{Z}}({\bf{E}};{\bf{g}})d{\bf{E}}}, \label{07}
\end{equation}
where -$\delta{U_{L_{x}}}({\bf{E}}|x,{\bf{g}})$  shows the
interaction potential energy  between  some steric spin chain with
energy of ${\bf{E}}$ and external field (\ref{02}), and
${\bf{Z}}({\bf{E}};{\bf{g}})$ denotes the energy distribution
function (partition function) of $1D$ steric spin chain
configurations. The definition of distribution function will be
given in section  4. Recall, that in (\ref{07}) we take into
account only negative values of ${\bf{E}}$, because only for these
values the spin chins maybe stable.

Taking the average value of expression (\ref{07}), we obtain the
following relation for the mean value of the potential increment:

$$
 -\langle\delta{U_{V}}(\vec{r};{\bf{g}})\rangle_{V}=-
\langle{\delta{U_{L_{x}}}({\bf{E}}|x,{\bf{g}})}\rangle_{(x,\updownarrow)}=
{\overline\alpha_{dip}}\langle{\vec{E}^{\,2}(x;{\bf{g}})}\rangle_{x},
$$

\begin{equation}
\langle{\delta{U_{L_{x}}}({\bf{E}}|x,{\bf{g}})}\rangle_{(x,\updownarrow)}=
\frac{1}{L_x}\int_0^{L_x}\langle{\delta{U_{L_{x}}}({{\bf{E}}|x,\bf{g}}
)}\rangle_{(\updownarrow)}d\,x, \label{08}
\end{equation}
where the bracket $\langle{...}\rangle_x$ means the integration
over $x$ on a scale $L_x$, the mean value of a full averaging
potential energy of interaction between the spin chains and the
external field  is defined by term
-$\langle{\delta{U_{L_{x}}}({\bf{E}}|x,{\bf{g}})}\rangle_{(x,{\updownarrow)}}$.

{\bf{Definition 1.}} In the relation (\ref{08}) parameter
${\overline{\alpha}_{ch}}$ will be named a collective polarization
coefficient of steric $1D$ spin chain. It is given by formula:

\begin{eqnarray}
\overline{\alpha}_{ch}=-\frac{ \langle{\delta{U_{L_{x}}}
({\bf{E}}|x,{\bf{g}})}\rangle_{(x,\updownarrow)}}
{\bigl\langle{\vec{E}^{\,2}(x;E_0,\Omega,\varphi_0)}\bigr\rangle_{x}}.
\label{09}
\end{eqnarray}
Note, that $\overline{\alpha}_{ch}$ is complex amount and
describes the average value of steric spin chain polarizability
with taking into account the lattice relaxation. When
$E_0\rightarrow 0$ it is simple to show that
$\overline\alpha_{ch}\rightarrow 0$.

Now one can write down the expression for the sum in the right
part of the relation (\ref{01}), which takes into account the spin
chains orientation effects in the external field:
\begin{eqnarray}
\sum_{m}{N_m}{\alpha_m}=\sum_m
{N^{0}_m}\alpha^{0}_{m}+{N_{ch}}{\overline\alpha}_{ch} =
\sum_m{N^{0}_m}\alpha^{0}_{m}-
{n_{dip}}N^{-1}_x\frac{\bigl\langle{\delta{U_{L_{x}}}
({\bf{E}}|x,{\bf{g}}})\bigr\rangle_{(x,\updownarrow)}}
{\bigl\langle{\vec{E}^2(x;E_0,\Omega,\varphi_0)}\bigr\rangle_{x}},
\label{10}
\end{eqnarray}
where  $N_{ch}=n_{dip}N_x^{-1}=(d_0(T))^{-3}N_x^{-1}$
-concentration of steric spin chains. From (\ref{10}) it follows
that in areas where the field strength is small the orientation
correction vanishes. In other words the external field on
wavelength scale (\ref{02}) creates the alternating
inhomogeneities with different dielectric constants. These layers
are stable in nanosecond scale $\Delta{t}\sim10^{-10}sec=0.1\,
ns$.

Using  (\ref{10}) we can generalize Clausius-Mossotti equation on
the space-time scale of standing wave taking into account
orientation effects:

\begin{equation}
\frac{\epsilon_{st}({\bf{g}})-1}{\epsilon_{st}({\bf{g}})+2}=\Lambda({\bf{g}}),
\quad where\quad \Lambda({\bf{g}})\simeq
\frac{4\pi}{3}\biggl[\sum_{m}N^{0}_{m}\alpha^{0}_{m}-
\frac{1}{d_0^3N_x}\cdot\frac{\langle{\delta{U_{L_{x}}}
({\bf{E}}|x,{\bf{g}}}\rangle_{(x,\updownarrow)}}
{\langle{\vec{E}^2(x;E_0,\Omega,\varphi_0)}\rangle_{x}}\biggr].
\label{11}
\end{equation}
 Note that $\epsilon_{st}$ is the label of stationary dielectric
constant.

 Now we turn to the study of the dispersion property of
frequency-depending dielectric constant (dielectric permittivity
function).

In the theory of dielectric relaxation, one writes the
frequency-dependent dielectric constant $\epsilon(\omega)$ with
Williams-Watts \cite{Williams} function of dielectric relaxation
by relation \cite{Montroll}:
\begin{equation}
\frac{\epsilon(\omega)-\epsilon_{\infty}}{\epsilon_{s}-\epsilon_{\infty}}
=\varrho(\sigma,\omega),\qquad\varrho(\sigma,\omega)=
-\int_{0}^{\infty}e^{-i\omega{t}}\Bigl[{d\,F_{\sigma}(t)}/{d\,t}\Bigr]dt,
\label{12}
\end{equation}
where $\epsilon_{\infty}=\epsilon(\omega\rightarrow\infty)$ is the
high-frequency limit of the dielectric constant and
$\epsilon_{s}=\epsilon(\omega\rightarrow\,0)$, the static
dielectric constant which can be defined from generalized
Clausius-Mossotti equation. In  (\ref{12}) the function $F(t)$
describes the decay of polarization of a dielectric sample with
time after sudden removal of steady polarizing electric field. The
frequency-dependent dielectric constants of broad class of
materials including polymeric systems and glasses may be
interpreted in terms of the Williams-Watts \cite{Williams}
polarization decay function:
\begin{equation}
F_{\sigma}(t)=exp[-(t/\tau)^{\sigma}],\qquad 0<\sigma<1,
\label{13}
\end{equation}
where exponent $\sigma$ and the time constant $\tau$ depending on
the material and fixed external conditions such as temperature $T$
and pressure.

The relation (\ref{12}) can be generalized for case of external
field as we are interested in the time scale lesser than the time
interval during which an external standing electrical wave may be
considered stationary. After substitution
$\epsilon(\omega)\rightarrow\epsilon(\omega,{\bf{g}})$ and
$\epsilon_{s} \rightarrow\epsilon_{st}({\bf{g}})$ the Eq.
(\ref{12}) we can transform and write in the form:
\begin{equation}
\frac{\epsilon(\omega,{\bf{g}})-\epsilon_{\infty}}{\epsilon_{st}
({\bf{g}})-\epsilon_{\infty}}
=\varrho(\sigma,\omega),\qquad\varrho(\sigma,\omega)
=-\sigma\int_{0}^{\infty}e^{-i\lambda{s}-s^{\sigma}}
s^{\sigma-1}ds,\quad \lambda=\omega{\tau},\quad s=t/\tau.
 \label{14}
 \end{equation}
For example in Debye's classical theory of dielectric relaxation
$\sigma=1$ \cite{Kit},  the integral in the right side of
(\ref{12}) has the form $(1+i\omega{\tau})^{-1}$.

\section{ Average interaction
potential between steric spin chain and external field }

As we have already shown in order to take into account the
contribution of orientation effects into the polarization one has
to calculate the total average value of the interaction potential
of steric spin chain with the external field
$-\langle{\delta{U_{L_{x}}}(E_0,\Omega,\varphi_0)}\rangle_{(L_x,{\updownarrow)}}$.

 Taking into account the fact that the external field is low
 i.e. $|\vec{E}(l_x-x)| \ll |\vec{E}_{loc}(l_x-x)|
 \cong  |\vec{E}_{int}(l_x-x)|$, we can apply the Taylor decomposition
 of dipole angular momentum:
\begin{eqnarray}
\vec{p}\,(l_x-x) \simeq
\vec{p}^{\,\,0}(l_x-x)+\delta\vec{p}\,(l_x-x), \quad \delta
\vec{p}\,(x) \sim \vec{E}(x;E_0,\Omega,\varphi_0), \label{15}
\end{eqnarray}
where $|{\delta\vec{p}(x)}| \ll |\vec{p}^{\,\,0}(x)|$, as well as
$\vec{E}^0_{int}(x-l_x)$  and $\vec{p}^{\,\,0}(x-l_x)$ are
correspondingly the field strength and the dipole angular momentum
of the molecule, located in the  $l_x$-th cell in the absence of
the external field. In case when the coordinate $x$ is outside of
$l_x$-th cell, the field  vanishes. Inside the cell they have
constant values. From the discussion above and taking into account
(\ref{07}), the relation (\ref{15}) can be represented in the
form:
\begin{equation}
-\delta U_{L_{x}}({\bf{E}}|x,{\bf{g}}))=
\sum_{l_{x}=0}^{L_{x}}\vec{p}^{\,\,0}
(l_x-x)\vec{E}(x;E_0,\Omega,\varphi_0)+
\sum_{l_{x}=0}^{L_{x}}\delta
\vec{p}\,(l_x-x)\vec{E}(x;E_0,\Omega,\varphi_0). \label{16}
\end{equation}
Using the fact that without an external field the spin system has
no polarization we conclude that first sum in (\ref{16}) vanishes.
In other words, the interaction potential has the following form:
\begin{equation}
-\delta U_{L_{x}}({\bf{E}}|x,{\bf{g}})=
\sum_{l_{x}=0}^{L_{x}}\delta
\vec{p}\,(l_x-x)\vec{E}(x;E_0,\Omega,\varphi_0). \label{17}
\end{equation}

Now we turn to the equation of motion for the steric spin-chain
with relaxation in $3D$ spin lattice in external field. Recall
that the interaction potential $-\delta
U_{L_{x}}({\bf{E}}|x,{\bf{g}})$ (see (\ref{07}) and (\ref{17}))
between $1D$ disordered spin chain and external field  does not
take into account relaxation with environmental spin chains.  It
is possible only after solution of the dynamical problem. The
resulting interaction potential in this case will be complex,
where the imaginary part characterizes the relaxation processes in
$3D$ lattice.

 In more general case those motions can have quantum characters and can be
represented by  complex Langevin-Schr\"odinger type stochastic
differential equation:
\begin{eqnarray}
\lambda\,\delta
U_{L_{x}}({\bf{E}}|x,{\bf{g}})=\lambda\,{\bf{E}}+\Psi^{-1}(d_t)^2
\Psi, \label{18}
\end{eqnarray}
where
$$
 t = x/d_0, \quad \lambda=2\mu/(\hbar^2d_0^2),\quad
 \mu=m_0/N^{1/(N-1)}, \quad (d_t)^2=d^2/dt^2,
$$

 $m_0$ and $\mu$ are the molecule (spin) mass and the spin
chain's effective mass correspondingly, $t$  denote of natural
parameter of evolution along the spin chain. In the equation
(\ref{18}) interaction potential $U_{L_{x}}({\bf{E}}|x,{\bf{g}})$
is a random complex function. Below we present its detailed
description.

Substituting
\begin{equation}
\Psi (t)= \exp \biggl(\int_0^{t}\Xi(t')dt'\biggr), \label{19}
\end{equation}
into (\ref{18}) and using the relations (\ref{19}), we obtain the
following  nonlinear complex SDE \cite{BGC}:
\begin{equation}
\Xi_t + \Xi^2 +\lambda\bigl({\bf{E}}-\overline
V\bigr)+\lambda\overline f(t)=0, \qquad\qquad \Xi(t)=\theta (t)+
i\vartheta (t), \quad \Xi_t= d\,\Xi/dt, \label{20}
\end{equation}
where
\begin{eqnarray}
\sum\limits_{l_x=0}^{L_x}\vec{p}\,(l_x-x)
\vec{E}\,(x;E_0,\Omega,\varphi_0)=\overline V +\overline
f(t),\qquad
 \overline V =\Bigl\langle{\sum_{{l_x}=0}^{L_x}\delta
\vec{p}\,(l_x-x)\vec{E}(x;E_0,\Omega,\varphi_0)}\Bigr\rangle_x.
\label{21}
\end{eqnarray}
In formulas (\ref{20}) and (\ref{21}) we have denoted by
$\overline V$ the mean value of the sum, and by $\overline f(t)$
its complex random part. Analyzing the contribution from different
mechanisms of molecule polarization in glass medium, we conclude
that under the influence of the external field (\ref{02}) with
frequency $\Omega \sim 10^9Hz $ the main part comes from elastic
dipole (dipole thermal polarization is not essential in this case
due to the large relaxation time $\tau\sim 10^{-4}\div
10^{-5}sec\,$\, \cite{Kit}, \cite{Berth}). Let us note that the
coefficient of elastic dipole polarization at low external fields
is determined by \,\cite{PKh}:
$$
\alpha_{dip}(l_x-x)=\Lambda^{-1}\Bigl( p^0\sin
[\beta(l_x-x)]\Bigr)^2=\frac {p^0}{E_{int}^0}\sin^2[\beta(l_x-x)],
$$
\begin{equation}
\delta\vec{p}\,(l_x-x)=\alpha_{dip}(l_x-x)\vec{E}\,(x;E_0,\Omega,\varphi_0),
\label{22}
\end{equation}
where $\beta(l_x-x)$ is the angle between the external
$\vec{E}\,(x;E_0,\Omega,\varphi_0)$  and the internal
$\vec{E}_{int}^{\,0}(l_x-x)$ fields,
$\Lambda({l_x})=\vec{p}^{\,\,0}(l_x)\vec{E}_{int}^{\,\,0}
(l_x)\cong{p^0}E^0_{int}$ is the dipole energy  in the field
$\vec{E}_{int}^{\,0}$. Following heuristic argumentation of Debby
\cite{Kit}, \cite{Williams,Montroll} one can write down the
expression for the elastic dipole polarization, which takes into
account the spin (polar molecule) relaxation process in the glass:
\begin{equation}
\Upsilon_{ch}(l_x-x)=\alpha_{dip}(l_x-x)/(1-i\Omega\tau)\cong
\frac{p^0\sin^2[\beta(l_x-x)]}{E_{int}^{\,0}(1-i\Omega\tau)},
\label{23}
\end{equation}
where $\tau$ is the spin relaxation time in the glass. It is very
small in the aforementioned media $\tau\sim 10^{-11}\div 10^{-15}$
sec (see, for example \cite{Kit}).

So, the equation (\ref{18}) with generalized coefficient of
elastic dipole polarizabilitiy (\ref{23})  will  describe the
motion of $1D$ random steric spin chain in the external field with
relaxation.

After substitution of (\ref{23}) in (\ref{21}) and simple
calculations we obtain:
\begin{equation}
\overline
V=\Bigl\langle{\sum\limits_{l_x=0}^{L_x}\Upsilon_{ch}(l_x-x){\vec{E}}^2
(x;E_0,\Omega,\varphi_0)}\Bigr\rangle_x=
-\frac{1+i\Omega\tau\,\,}{1+(\Omega\tau)^2}\frac{N_x
p^0\overline{E}_0^{\,2}}{4d_0^{\,3}
E_{int}^{\,0}}\left(1-\frac{\sin(2kL_x)}{2kL_x}\right), \label{24}
\end{equation}
where $\overline{E}_0(\varphi_0)=2E_0\sin{\varphi_0}$.

 Let us now investigate the properties of the random
function $\overline{f}(t)$. From the relations (\ref{21}) and
(\ref{23}) it is easy to find the random strength:
$$
\overline{f}(t)=-\frac{1+i\Omega\tau}{1+(\Omega\tau)^2}
\cdot\frac{p^0\overline{E}_0^2 }{4d_{\,0}^3
E_{int}^{\,0}}\xi(t),\qquad \xi(t)=\Bigl(1+\cos(2k_tt)\Bigr)
\sum\limits_{l_x=0}^{L_x}\cos2\overline\beta(l_t-t)
$$
\begin{equation}
\overline\beta(l_t-t)=\beta(l_x-x),\qquad
k_t=\hbar^{-1}\sqrt{2\mu}\,k,\qquad l_t=\hbar{l}/\sqrt{2\mu}.
\label{25}
\end{equation}
If the phase $\beta$  is homogeneously distributed along the
interval $[0,\pi]$, then for the mean value we obtain:
\begin{equation}
M\overline{f}(t)=\langle{\overline{f}(t)}\rangle=0,\quad
M\xi(t)=\langle\xi(t)\rangle=0. \label{26}
\end{equation}
For the autocorrelation function one can write the following
expression (see, for example \cite{GTT}):
\begin{equation}
R_{ff}(\overline{t})=\langle{\overline{f}(t)\overline{f}(t')}
\rangle=\frac{1}{2}\left(\frac {1+i\Omega\tau} {1+(\Omega\tau)^2}
\right)^2\left(\frac{N_xp^{\,0}\overline{E}_{\,0}^2}
{4d_0^{\,3}E_{int}^{\,0}}\right)^2\left(M\xi^{2}(\overline{t})\right),
\label{27}
\end{equation}
where $\overline{t}=t-t'$. Substituting (\ref{25}) in (\ref{27})
and carrying out straightforward calculations, we may obtain:
\begin{equation}
R_{ff}(\overline{t})=\langle{\overline{f}(t)\overline{f}(t')}
\rangle=2\overline{D}\delta(t-t'),\qquad \overline{D}=\frac{1}{4}
\left(\frac{1+i\Omega\tau}{1+(\Omega\tau)^2}
\right)^2\left(\frac{N_xp^0\overline{E}_0^2}{4d_0^{\,3}
E_{int}^0}\right)^{2}. \label{28}
\end{equation}
So, we have shown that the dispersion equation (\ref{16}), which
describes the relation between the external field and the local
polarization in disordered system, is reduced to the investigation
of the Langevin type nonlinear complex stochastic SDE with the
autocorrelation function (\ref{26})-(\ref{28}) of white noise. For
further investigation, it is convenient to represent the complex
equation (\ref{20}) as a system of two real equations:
\begin{equation}
\dot{\theta}+\theta^2-\vartheta^2+\lambda\bigl({\bf{E}}-V_r+f_r(t)\bigr)=0,
\label{29}
\end{equation}
\begin{equation}
\dot{\vartheta}+2\vartheta\theta+\lambda\bigl(V_i+f_i(t)\bigr)=0,
\label{30}
\end{equation}
where $\dot{\theta}=d_t\theta$, $\dot{\vartheta}=d_t\vartheta$,
$V_r=Re\overline{V}$, $V_i=Im\overline{V}$,
$f_r(t)=Re\overline{f}(t)$, and $f_i(t)=Im \overline{f}(t)$.

Now the problem is to find of evolution equation for the
conditional probability:
\begin{equation}
Q\bigl(\theta,\vartheta,t|\theta_0,\vartheta_0,t_0\bigr)=\Bigl\langle\delta
\bigl(\theta(t)-\theta(t_0))\delta(\vartheta(t)-\vartheta(t_0)\bigr)\Bigr\rangle
\biggl|_{\{\theta_0=\theta(t_0); \vartheta_0=\vartheta(t_0)\}},
\label{31}
\end{equation}
describing the probability  of fact that, the trajectory
$\bigl(\theta_0\equiv\theta(t_0),
\vartheta_0\equiv\vartheta(t_0)\bigr)$ at the initial moment
leaving of natural parameter $t_0$ from the point
$(\theta_0,\vartheta_0)$, at the arbitrary moment of $t$ it will
be turned  out to the suburb of point $(\theta,\vartheta)$.
Subject to SDE system (\ref{29}), (\ref{30}) the Fokker-Plank
equation can be easily found   \cite{Klyatskin} (see also
\cite{BGC}):
\begin{equation}
\frac{\partial{Q}}{\partial{t}}=D_r\frac{\partial^2{Q}}
{\partial\theta^2}+D_i\frac{\partial^2{Q}} {\partial\vartheta^2}+
\bigl(\theta^2-\vartheta^2+\lambda\bigl({\bf{E}}-V_r)\bigr)
\frac{\partial{Q}}{\partial{\theta}}+(2\theta\vartheta
+\lambda{V_i)}\frac{\partial{Q}}{\partial{\vartheta}}+4\theta Q,
\label{32}
\end{equation}
where $Q\equiv{{Q(\bf{E}}|\theta,\vartheta;t)}$, $D_r=\lambda^2Re
\overline{D}$ and $D_i=\lambda^2Im \overline{D}$. Note, that the
solution of  equation (\ref{32}) must  satisfy  the initial
condition:
\begin{equation}
Q({\bf{E}}|\theta,\vartheta;t)\biggl|_{t=t_0}=\delta(\theta-\theta_0)
\delta(\vartheta-\vartheta_0), \label{33}
\end{equation}
where the initial phases $\theta_0$ and $\vartheta_0$ are equal to
zero. Later we will be more interested in the  fixed  limit  of
the solution (\ref{32}), which is obviously received under values
$t\gg\bigtriangleup{t}=O(1)$ which is equivalent to the condition
$t\to{\infty}$. In that limit the equation (\ref{32}) is
simplified and accurate to the value
$O(\bigtriangleup{t}/t_{N_x})\ll O(1)$ may take the following
form:
\begin{equation}
D_r\frac{\partial^2{Q_s}}{\partial\theta^2}+D_i\frac{\partial^2{Q_s}}
{\partial\vartheta^2}
Q_s+\bigl(\theta^2-\vartheta^2+\lambda({\bf{E}}-V_r)\bigr)
\frac{\partial{Q_s}}{\partial{\theta}}
+\bigl(2\theta\vartheta+\lambda{V_i}\bigr)\frac{\partial{Q_s}}{\partial\vartheta}
+4\theta Q_s=0, \label{34}
\end{equation}
where $Q_s\equiv
Q_s({\bf{E}}|\theta,\vartheta)\equiv\lim_{t\to\infty}Q(\theta,\vartheta,t)$
and $t_{N_x}=N_x$.

 The (\ref{34}) is an elliptic type
differential equation for which there are not real
characteristics. This implies that the type of quasilinear
equation  depends on  which solution is considered and can be
different for different solutions. The solution of the equation
(\ref{34}) must satisfy the conditions:
\begin{equation}
Q_s\Bigl|_{S}=
\frac{\partial{Q_s}}{\partial{\textbf{n}}}\Bigl|_{S}=0,\quad\quad
|\textbf{n}|={(\theta^2+\vartheta^2)}^{1/2}, \label{35}
\end{equation}
where the border condition is set on a curve $S$ and $\textbf{n}$
is correspondingly the normal of curve.

Now one can write the value of the average interaction potential:
\begin{eqnarray}
\langle{\delta U_{L_x}({\bf{E}}|x,{\bf{g}})}\rangle_x={\bf{E}}+
\frac{1}{\lambda}\cdot\lim\limits_{t\to
t_{N_x}}\frac{1}{t}\int_0^{t_{N_x}}
\left(\dot{\theta}+i\dot{\vartheta}+{\theta}^2-{\vartheta}^2-
i2\theta\vartheta\right)dt\nonumber\\
={\bf{E}}+\frac{1}{\lambda}\cdot\lim\limits_{t\to
+\infty}\frac{1}{t}\int_0^{+\infty}
\left(\dot{\theta}+i\dot{\vartheta}+{\theta}^2-{\vartheta}^2-
i2\theta\vartheta\right)\,dt
+O\left(\frac{\Delta{t}}{t_{N_x}}\right).
 \label{36}
\end{eqnarray}
Using the Birgoff ergotic  hypothesis \cite{LGP} we can change
integration by natural parameter $t$ to integration by stationary
distribution $Q_s({\bf{E}}|\theta,\vartheta)$ and finally find the
following expression for mean value of the potential energy
increment:
\begin{eqnarray}
\langle{\delta U_{L_x}({\bf{E}}|x,{\bf{g}})}\rangle_x={\bf{E}}+
\frac{1}{\lambda{R}}\int_{-\infty}^{+\infty}\bigl[\theta^2-
\vartheta^{2}-i2\theta\vartheta\bigr]Q_s({\bf{E}}|\theta,\vartheta)
\,d\theta\,d\vartheta  +O(N_{x}^{-1}),
 \label{37}
\end{eqnarray}
where $R=\int_{-\infty}^{+\infty}Q_s({\bf{E}}|\theta,\vartheta)
{d\theta}{d\vartheta }$ denoted normalization constant.

\section{Statistical mechanics of the steric $1D$ random spin
chain }

Before turning to the effective statical dielectric constant
calculation the two important problems have to be solved. The
first one is the calculation of the partition function
${\bf{Z}}({\bf{E}};{\bf{g}})$ for the steric $1D$ random spin
chain with energy ${\bf{E}}$, whereas the second one is the
determination of the thickness of layers on a scale of this chain
(i.e. the fragmentation  of the media on layers with different
dielectric constant).

First, let us consider the statistical properties of the $1D$
steric spin chain. As it is shown, the type of a compound
$(a-SiO_2)$ can be used as a canonical model for such network
glasses describing continuous random networks of atom and bonds
\cite{WHZ},\cite{Tu} (see Fig 1). It means that  Heisenberg
spin-glass Hamiltonian \cite{Bind} can be suitable for describing
of $3D$ disordered spin system. Note that this allows to use the
ergodic hypothesis (see (\ref{07})) and  make reduction of $3D$
spin lattice dynamical problem to the problem of $1D$ complex
Langevin-Schr\"odinger type stochastic differential equation (see
(\ref{18}), (\ref{20})) and the problem of construction
statistical mechanics of steric disordered spin chain system. As
 it is easy to see, this type of reduction in a certain sense is
conditional because  both problems are interrelated.

For further investigations it is favorable to consider the
spin-glass Hamiltonian of type:

\begin{eqnarray}
{\bf{H}}(N_x;{\bf{g}})=-\sum_{ij}^{N_x}
J_{ij}(r_{ij})\vec{S}_{i}\vec{S_j}+p^0\sum_{i}^{N_x}
\vec{E}_i\vec{S_i},\qquad\,\,
\vec{S_i}\equiv\frac{\vec{p}_i^{\,\,0}}{p^{\,0}}, \label{38}
\end{eqnarray}
where $r_{ij}=|i-j|d_0(T)+\eta_{ij}$ is a distance between $1D$
spins $\vec{S}_{i}$ and $\vec{S_j}$ (classical vectors of unit
length), $\eta_{ij}$ is its random part, subject to Gaussian
distribution with zero mean value and unit variance, $J_{ij}$ is
nearest-neighbor interaction constant, which depends on the
distance between spins. It may be positive or negative, the
external field $\vec{E}_i$ is defined by means of formulas
(\ref{02}).

Now the problem is the construction of the energy distribution
function of spin-chains ${\bf{Z}}({\bf{E}};{\bf{g}})$. However,
the problem is that the time scale, on which we make statistical
study of the system, is very short ($\lesssim
10^{-10}sec=0,1nsec$), while the characteristic thermal relaxation
time in amorphous media \cite{Kit} is of order
${\Omega_T^{-1}}\sim10^{-4}\div10^{-5}sec$, where $\Omega_T$  is
the frequency of thermal fluctuations. The last fact means that
the temperature and related thermodynamical constructions become
meaningless in our problem. Nevertheless, some structural
similarity between gas and amorphous media is evident. At that a
steric spin chain will correspond to an atom in gas. Since in the
equilibrium state the average energy value per atom in the gas is
$\frac{3}{2}kT$, a corresponding value in this case will be the
chain's energy in the equilibrium state (chain's energy without
external field). However, the considered system has a specific
peculiarity. The point is that the equilibrium state in gas is
characterized by one temperature whereas the spin system can be in
the equilibrium state in any negative energy. These energies
coincide with the local minimums of non-perturbed Hamiltonian and
as  it is well-known their numbers can be  high
\cite{Bind,ChinKunHu}. In other words, in this case the phase
space can be decomposed uniquely into micro-canonical states
associated with different thermodynamic equilibrium states
\cite{Hemm}.

The Hamiltonian (\ref{38}) in case of the  external field absence
can be rewritten in spherical coordinate system as follows:
\begin{equation}
{\bf{H}_0}(d_0(T),N_x)=\sum_{\{ij\}=1,\,\,i\neq{j}}^{N_x}
J_{ij}(r_{ij})\Bigl(\cos\psi_i\cos\psi_j\cos(\phi_i-\phi_j)
+\sin\psi_i\sin\psi_j\Bigr), \label{39}
\end{equation}
For the determination of local minimums one has to solve the
following algebraic equations:
$$
\Phi_{\psi_i}({\bf{\Theta}})=\frac{\partial{\bf{H}}}{\,\,\partial\psi_i}=
\sum_{j=1}^{N_x}J_{ij}(r_{ij})\Bigl(-\sin\psi_i\cos\psi_j\cos(\phi_i-\phi_j)
+\cos\psi_i\sin\psi_j\Bigr)=0,
$$
$$
\Phi_{\phi_i}({\bf{\Theta}})=\frac{\partial{\bf{H}}}{\,\,\partial\phi_i}=-
\sum_{j=1}^{N_x}J_{ij}(r_{ij})\cos\psi_i\cos\psi_j\sin(\phi_i-\phi_j)=0,
$$
\begin{equation}
\Phi_{r_{ij}}({\bf{\Theta}})=\frac{\partial{\bf{H}}}{\,\,\partial{r_{ij}}}=
\sum_{j=1}^{N_x}\frac{\partial J_{ij}}{\partial
r_{ij}}\Bigl(\cos\psi_i\cos\psi_j\cos(\phi_i-\phi_j)+\sin\psi_i\sin\psi_j\Bigr)=0,
 \label{40}
\end{equation}
where ${\Theta}_i=(\psi_i,\phi_i)$ are angles of $i$-th spin
($\psi_i$ is the polar and $\phi_i$ azimuthal angles),
${\bf{\Theta}}=({\Theta_1},{\Theta_2}....{\Theta_{N_x}})$
correspondingly describe the angular part of spin configuration.
Now suppose that non-perturbed Hamiltonian for fixed averaged
distance between spins $d_0(T)$ has $n$ local minimums of function
${{\bf{E}}(d_0(T))}$, each corresponds to $M_j$ spin
configurations $\{{\bf{\Theta}}^{0i}\}$, where $i=0,1...M_\nu$.
The number of all configurations, which correspond to local
equilibrium states $M_{full}=\sum_{j=1}^{N}M_\nu$.
Correspondingly, the weight of every equilibrium state may be
defined by formulas:
\begin{equation}
P_\nu({\bf{E}}_\nu;d_0(T))=M_\nu/M_{full},\qquad
\sum_{\nu=1}^{N}P_\nu({\bf{E}}_\nu;d_0(T))=1. \label{41}
\end{equation}

Thus we can propose to use the following statistical weight
instead of the canonical distribution at multi-equilibrium state:
\begin{equation}
W_\nu({\bf{H}};{\bf{g}})=P_\nu({\bf{E}}_\nu;d_0(T))\exp
\biggl\{-\frac{{\bf{H}}(N_x;{\bf{g}})}
{{\bf{E}}_\nu(d_0(T))}\biggr\}, \label{42}
\end{equation}
where ${\bf{E}}_\nu(d_0(T))$ is the energy of spin chain in the
absence of an external field. Recall, that the multicanonical
ensemble was introduced in the \cite{BN} as an approach to
simulate a strong first-order phase transitions. It has been very
successful in sampling a large barrier.

 Now, taking into account (\ref{39})-(\ref{42}), one can
present the expression of the energy  distribution function  for
$N_x$ spin system (this expression can be explained as a local
partition function in above mentioned multicanonical
thermodynamics):
\begin{equation}
{\bf{Z}}_\nu({\bf{E}}_\nu;{\bf{g}},\textbf{q})=
P\bigl({\bf{E}}_\nu;d_0(T)\bigr)\int{\frac{d\Omega_1}{4\pi}}\cdot\cdot\cdot
{\frac{d\Omega_{N_x}}{4\pi}}
\exp\biggl\{-\frac{{\bf{H}}(N_x;{\bf{g}})}{{\bf{E}}_\nu
(d_0(T))}\biggr\}, \label{43}
\end{equation}
where  $\textbf{q}$ describes the  set of random distances
$\eta_{ij}$ and random angles $\beta_i$, correspondingly
$d\Omega_i$ is an element of the solid angle $\Omega_i$ containing
the unit vector $\vec{S}_i$.

Energy distribution function of spin chain can be essentially
simplified after the suggestion that in Hamiltonian (\ref{38})
only nearest-neighboring spins interact, i.e. $J_{ij}\equiv0$ if
$|j-i|>2$. In this case the multidimensional integral can be taken
exactly as follows. The integration starts from the end of the
chain. When integrating over $d\Omega_i$ we take as a polar axis
the direction of the vector $(J_{i(i-1)}\vec{S}_i+p^0\vec{E}_i)$.
Then it is easy to obtain the following expression:
$$
{\bf{Z}}_\nu({\bf{E}}_\nu;{\bf
g},\textbf{q})=P_\nu\prod_{i=1}^{N_x}
\biggl[\frac{1}{2}\int_0^{\pi}
\exp\{K_{i;\nu}\cos\psi\}\sin\psi\,d\psi\biggr]
=P_\nu\bigl({\bf{E}}_\nu;d_0(T)\bigr)
\prod_{i=1}^{N_x}\frac{\sinh{K_{i;\nu}}}{K_{i;\nu}},
$$
\begin{equation}
K_{i;\nu}=\frac{1}{{\bf{E}}_\nu}\Bigl[J_{(i-1)i}^2+2p^0E_iJ_{(i-1)i}
\cos\beta_i+(p^0)^2E_i^2\Bigr], \label{44}
\end{equation}
where $\beta_i$  is the random angle between the vectors
$\vec{S}_{i-1}$  and $\vec{E_i}$. Supposing that the distribution
of spin $\vec{S}_{i}$ around field $\vec{E_i}$ direction is
isotropic, one can perform an integration by angle $\beta_i$.
After simple integration with the help of formula
\cite{Abramowitz}:
$$
Ei(ax)=\int\frac{e^{ax}}{x}dx=\ln|x|+\sum_{k=1}^{\infty}
\frac{\bigl(ax\bigr)^k}{k!k}, \qquad a\neq{0},
$$
 we can find:
\begin{eqnarray}
{\bf{Z}}({\bf{E}};{\bf g},\eta)=P\bigl({\bf{E}};d_0(T)\bigr)
\prod_{i=1}^{N_x}\biggl[\frac{1}{2}\int_0^{\pi}
\frac{\sinh{K_{i}}}{K_{i}}\sin\beta_i\,d\beta_i\biggr]\qquad\qquad\qquad
\nonumber\\
=P\bigl({\bf{E}};d_0(T)\bigr)
\prod_{i=1}^{N_x}\frac{1}{4a_{i;\nu}}\biggl\{\Bigl
[Ei(b_{i}+a_{i})- Ei(-b_{i}-a_{i})\Bigr]-\Bigl
[Ei(b_{i}-a_{i})- Ei(-b_{i}+a_{i})\Bigr]\biggr\},\nonumber\\
 a_{i}=  \frac{1}{{\bf{E}}}
2p^0E_iJ_{(i-1)i},\qquad
b_{i}=\frac{1}{{\bf{E}}}\bigl[J_{(i-1)i}^2+(p^0)^2E_i^2\bigr].
\qquad\qquad\qquad \label{45}
\end{eqnarray}
For simplification of expression (\ref{45}) and below will be
admit index $\nu$. In the (\ref{45}) symbol $\eta$ describes the
set of random distances, $Ei(x)$ is the exponential integral
function. It is simple that in spite of the fact that every member
in parentheses near the points $E_c^+=J_0/p^0$ or $E_c^-=-J_0/p^0$
has a singularity nevertheless their sum  is an analytical
function.

 In the paper \cite{EdwAnd} at the first time it was suggested
that one may describe spin glass by a Hamiltonian of the type
(\ref{38}), where spins are put onto the sites of a regular
lattice, and disorder is introduced  by a suitable distribution
$W(J_{(i-1)i})$ of exchange bonds. A standard choice is the
Gaussian Edwards-Anderson model \cite{EdwAnd} (see also
\cite{Sherrington}):
\begin{eqnarray}
W(J_{(i-1)i})=\frac{1}{\sqrt{2\pi(\Delta{J_{(i-1)i}})^2}}\exp
\biggl\{-\frac{\bigl(J_{(i-1)i}-J_0\bigr)^2}
{2(\Delta{J_{(i-1)i}})^2}\biggr\}\qquad
\nonumber\\
J_0=\bigl<J_{(i-1)i}\bigr>_{av},\quad
\bigl(\Delta{J_{(i-1)i}}\bigr)^2=\bigl<J_{(i-1)i}^2\bigr>_{av}
-\bigl<J_{(i-1)i}\bigr>_{av}^2.
 \label{46}
\end{eqnarray}
Recall that for this model $J_0$ and $\Delta{J_{(i-1)i}}$ are
independent on distance and scaled with spin number $N_x$ as
\begin{eqnarray}
\bigl<J_{(i-1)i}\bigr>_{av}=J_0\propto{N_x^{-1}},\qquad
\Delta{J_{(i-1)i}}\propto{N_x^{-1/2}},
 \label{47}
\end{eqnarray}
to ensure a sensible thermodynamic limit.  Eq.s (\ref{46}) and
(\ref{47})  $\bigl<...\bigr>_{av}$ describe the averaging
procedure. Now we can conduct averaging of function (\ref{45}) by
distribution (\ref{46}) and find the local partition function near
some equilibrium energy $\textbf{E}_\nu$ of disordered spin chain:
\begin{eqnarray}
{\bf{Z}}({\bf{E}};{\bf g})=P\bigl({\bf{E}};d_0(T)\bigr)
\bigl<{\bf{Z}}({\bf{E}},J;{\bf g})\bigr>_J,\qquad\qquad\qquad
\nonumber\\
\bigl<{\bf{Z}}({\bf{E}},J;{\bf g})\bigr>_J= \prod_{i=1}^{N_x} \int
dJ_{(i-1)i}W\bigl(J_{(i-1)i}\bigr)G({\bf{E}},J_i;{\bf
g}),\qquad\nonumber\\
 G({\bf{E}},J_i;{\bf
g})=\frac{1}{2a_{i}}\biggl\{\Bigl [Ei(b_{i}+a_{i})-
Ei(-b_{i}-a_{i})\Bigr]-\Bigl [Ei(b_{i}-a_{i})-
Ei(-b_{i}+a_{i})\Bigr]\biggr\}.
 \label{48}
\end{eqnarray}
 Taking into account the fact that on the scale of half-wavelength
$N_x\gg{1}$ as well as that $G({\bf{E}},J_i;{\bf g})$ is an
analytical function for computation of integrals in partial
partition function (\ref{48}), the Laplace asymptotic method
\cite{Fedoryuk} may be used:
\begin{eqnarray}
{\bf{Z}}({\bf{E}},{\bf g})\approx{P\bigl({\bf{E}};d_0(T))}
\prod_{i=1}^{N_x}
 \frac{1}{4a_{i}^0}\biggl\{\Bigl[Ei(b_{i}^0+a_{i}^0)-
 Ei(-b_{i}^0-a_{i}^0)\Bigr]-\qquad\quad\qquad\quad
 \nonumber\\
 \Bigl[Ei(b_{i}^0-a_{i}^0)-
Ei(-b_{i}^0+a_{i}^0)\Bigr]\biggr\}, \qquad\quad
 a_{i}^0=\frac{1}{{\bf{E}}} 2p^0E_iJ_0,\quad
b_{i}^0=\frac{1}{{\bf{E}}}\bigl[J_0^2+(p^0)^2E_i^2\bigr].
\qquad\qquad
 \label{49}
\end{eqnarray}
It is simple to show that in the limit of $E_i\to{0}$ the
correspondence ${\bf{Z}}({\bf{E}},{\bf g})\to
P\bigl({\bf{E}};d_0(T)\bigr)$ takes place.

Note, that averaging procedure in the framework of multicanonical
ensemble is carried  by the following formula:
\begin{equation}
\langle{Y(x,{\bf{g}})}\rangle_{(\updownarrow)}=\lim_{N\to\infty}
\frac{\sum_{\nu=1}^NY({\bf{E}}_\nu|x,{\bf{g}})
{\bf{Z}}({\bf{E}}_\nu;{\bf{g}})}{\sum_{\nu=1}^N{\bf{Z}}({\bf{E}}_\nu;{\bf{g}})}=
\frac{\int_{-\infty}^0 {Y({\bf{E}}|x,{\bf{g}})}
{\bf{Z}}({\bf{E}};{\bf{g}}) d{\bf{E}}} {\int_{-\infty}^0
{\bf{Z}}({\bf{E}};{\bf{g}})d{\bf{E}}}, \label{50}
\end{equation}
where $Y(x,{\bf{g}})$ is an averaged physical parameter.

 For $1D$ random steric spin chain system in analogy to usual thermodynamics
Helmholtz type free energy may be specified near the local
equilibrium state with energy ${\bf{E}}_\nu$. In this case the
free energy amount with the expectation of one spin and the per
unit of equilibrium  state energy $\textbf{E}_\nu$ is defined:
\begin{eqnarray}
\textbf{F}({\bf{E}};{\bf
g})=-\frac{1}{N_x}\ln{\bf{Z}}({\bf{E}};{\bf g}).
 \label{51}
\end{eqnarray}
All macroscopic thermodynamic properties of $1D$ random steric
spin chain can be obtained by free energy derivatives. The simple
investigation of expression for free energy  shows that in the
steric $1D$ random spin chain model  with the nearest-neighbor
interaction  depending on amplitude of external field $E_0$ there
are not phase transition phenomena. However in this case under the
even low fields $E_i\sim 1/N_x$ the free energy of spin system
essentially changes (see Fig 6).
\begin{figure}[h]
\begin{center}\includegraphics[height=70mm]{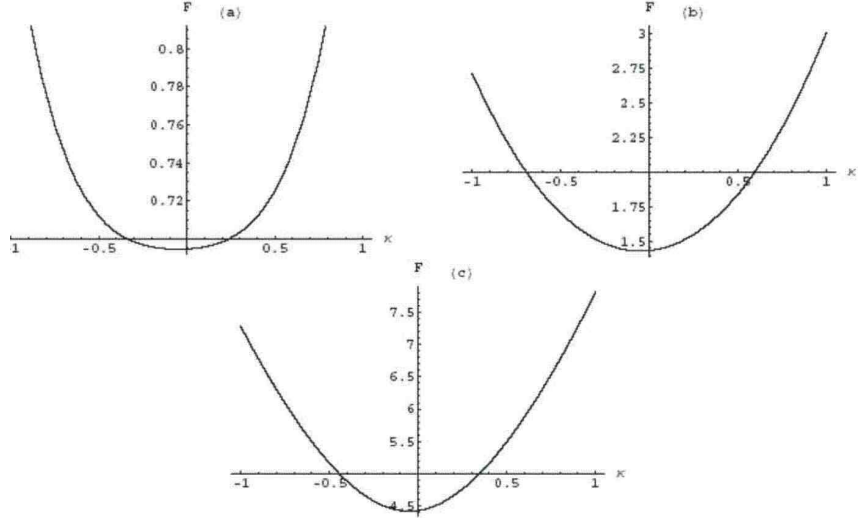}\end{center}
 \caption{The
Helmholtz free energy for any average energy
 of spin chain $\textbf{E}_\nu$ $\bigl($see figures (a),(b) and (c)
 $\bigr)$ and amplitude of external field $E_0$. On the figures
 axis $k$ shows  the parameter which is proportional to
  external fields amplitude $E_0$.}
\end{figure}

Particularly, the order parameter for disordered spin-glass medium
is described by $\sum_ip_i^{\,x}$ and without the external field
on the scale of the period of standing wave it is equal to zero,
where $p_i^{\,x}$ is a projection of spin on the direction of
external field propagation. It becomes non-zero at the weak
external field $E_i\sim |E_c^\pm|$ due to a symmetry breaking. In
this case nanoparticles (spin chains) with the \emph{super spins}
(macropolarizations) are generated on the microscale space and
nanoscale time.

\section{Dielectric permittivity  of the neighboring layers}

Now we can examine the question of dividing the dielectric medium
on the regions with extremely different polarizations. Recall that
the polarized medium without an external field   has a zero
macroscopic polarization on the space scale
$10^{-4}\div\,10^{-5}cm$. After turning on the external field in
the medium (on the scale of wavelength of standing electrical
field) the amplification of the orientational effects and, as a
result, the initiation of  macroscopic polarization is possible.
Assuming that the collective orientation effects, generated in
those regions of media, where voltage of external field $E$ is of
order to the critical value $ E_c$, we can divide the scale of
wavelength of external field on the forth region. In the first and
third regions the macroscopic polarization of medium is zero
whereas in the second and forth regions  it is different from zero
(see Fig. 7). Note that the height of these layers may be computed
with the help of numerical experiments.

\begin{figure}[h]
\begin{center}\includegraphics[height=60mm]{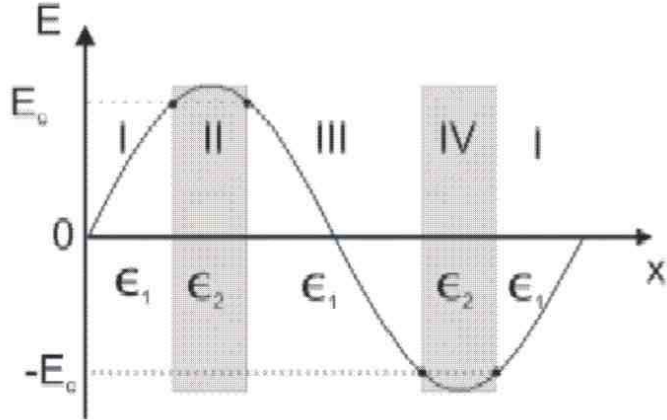}\end{center}
\caption{ \emph{The regions with different polarizations and
correspondingly different dielectric constants on the one
wavelength $\lambda_s$ scale}. }
\end{figure}

Now, with calculation of effective dielectric constant at the
half-wavelength scale of external electromagnetic field at the
time intervals $\Delta{t}\ll{2\pi/\Omega}$. Using (\ref{11}), we
can obtain:
\begin{equation}
\epsilon_{st}({\bf{g}})=\frac{1+2\Lambda({\bf{g}})}
{1-\Lambda({\bf{g}})\,}. \label{52}
\end{equation}
Remember, that here we have taken into account the contribution
from the elastic dipole polarization only. It is easy to see that
the nominator of the expression (\ref{52}) has no singularities
and is bounded by the imaginary part of the potential increment
(\ref{37}). The important characteristic is the dielectric
constant's difference in neighboring layers. Taking into account
the relations (\ref{11}) and (\ref{52}) one can get:
\begin{equation}
\delta\epsilon_{st}({\bf{g}})=\epsilon_{st}({\bf{g}})
-\epsilon_{st}(0)=(1-\epsilon_{st}(0))+\frac{3\Lambda({\bf{g}})}
{1-\Lambda({\bf{g}})}.
 \label{53}
\end{equation}
Dielectric permittivity of layer  in the external field can be
calculated simply:
\begin{equation}
\epsilon_{st}({\bf{g}},\omega)=\epsilon_{\infty}
+\bigl(\epsilon_{st}(\textbf{g})-\epsilon_{\infty}\bigr)\varrho(\sigma,\omega).
 \label{54}
\end{equation}
From equation (\ref{54}) taking into account Debye relaxation
model in the $x$-rays region we find:
\begin{equation}
\epsilon_{st}({\bf{g}},\omega)\simeq 1
-\bigl(\epsilon_{st}(\textbf{g})-1\bigr)\biggl\{i\frac{\omega_p}{\omega}
+\Bigl(\frac{\omega_p}{\omega}\Bigr)^2\biggr\},
 \label{55}
\end{equation}
where $\omega_p$ is the plasma frequency of  media.

 Now  we can  write the expression for  difference of frequency-dependent
dielectric constants (dielectric permittivity function) in
neighboring layers:
\begin{equation}
\delta\epsilon({\bf{g}},\omega)=\epsilon_{st}({\bf{g}},\omega)-
\epsilon_{st}(\omega)=\delta\epsilon_{st}({\bf{g}})\,\varrho(\sigma,\omega),
\label{56}
\end{equation}
which in the $x$-rays region has a form:
\begin{equation}
\delta\epsilon({\bf{g}},\omega)\simeq-\delta\epsilon_{st}
({\bf{g}})\,\biggl\{i\frac{\omega_p}{\omega}
+\Bigl(\frac{\omega_p}{\omega}\Bigr)^2\biggr\}. \label{57}
\end{equation}
So, we  constructed the dielectric permittivity function
(\ref{54})-(\ref{55}) for spin-glass medium under the  external
standing electromagnetic wave (\ref{02}), which is  correct on the
micrometer space scale and nanosecond time scale.

Note, that the main peculiarity of expression (\ref{54}) concludes
in the fact that  $\epsilon_ {st}({\bf{g}})$ is a complex
function. Moreover depending on parameters of weak external field
$\epsilon_ {st}({\bf{g}})$ can have big imaginary value. This fact
in contrast to usual case builds the media with radically new
properties.  Recall when the external field is absent the
dielectric permittivity function depends on $\epsilon_{s}$, which
is order of unit real positive constant.

\section{Conclusion}
In this article a new microscopic approach is developed for the
study of the properties of stationary dielectric constant and
dielectric permittivity function in the dielectric medium under
the external standing electromagnetic field. The approach consists
of the following two general steps:
\begin{enumerate}
\item Generalization of Clausius-Mossotti equation for dielectric
constant in the external standing electromagnetic wave; \item
generalization of the equation for dielectric permittivity
function taking into account the previous results.
\end{enumerate}
Mathematically the problem is solved as follows.  The dielectric
medium in the external electromagnetic field is modelled as a $3D$
spin-glass system under the external field. Note that all general
changes of properties of media take place in the scale of
wavelength of external field. For that the layer of medium which
consist from disordered $1D$ spin chains with length of order of
wavelength in detail has been investigated. Taking into account
the fact that the distribution of spin chains in the infinite
$(x,y)$ plan is isotropic we can use the Birgoff ergodic
hypothesis (see (\ref{06})-(\ref{07})) and conditionally reduce
the initial $3D$ spin-glass problem to two $1D$ problems. It means
that we can investigate two $1D$ problems separately only in this
case the parameters of the first $1D$ problem must be taken into
account during the solution of the second $1D$ problem.

 In the work
all formal definitions for accounting the investment of
orientational effects in the stationary and frequency-depended
dielectric constants computation are adduced.

The first $1D$ problem is related to one-dimensional disordered
quantum $N$-particle system with relaxation.  The investigation of
the motion in this system takes place in the framework of complex
Langevin-Schr\"{o}dinger type SDE (\ref{18}), which can be
transformed to  $2D$ system of nonlinear Langevin type SDE
(\ref{29})-(\ref{30}).
 For probability distribution of
interaction potential in $1D$ spin chain with certain energies
$\textbf{E}$ and external standing electromagnetic wave
Fokker-Plank equation (\ref{32}) is obtained, which is found after
using white noise  model for stochastic forces and system of SDE
(\ref{29})-(\ref{30}). In the limit of long-range distance
$t\to\infty$, the probability distribution
$Q(\textbf{E}|\theta,\vartheta;t)$ tends to the limit of its
stationary limit $Q_s(\textbf{E}|\theta,\vartheta)$ which
satisfies the elliptic type differential equation (\ref{34}).
Note, that this type of differential equation (\ref{34}) depending
on parameters can have extremely various solutions between which
there aren't smooth transitions. In another words this type of
behavior in the solutions we can interpret as a critical
phenomenon in the $3D$ spin-glass system.
\begin{figure}[h]
\begin{center}\includegraphics[height=70mm]{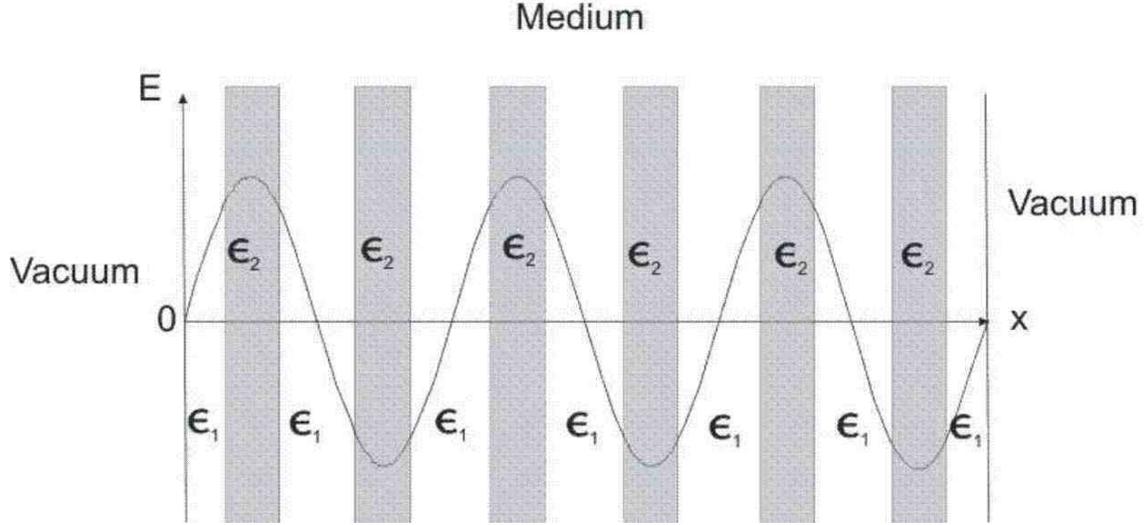}\end{center}
\caption{ \emph{The regions of different polarizations}. }
\end{figure}

 The second $1D$ problem includes the computation of $1D$
 random spin chain energy  distribution in the
 external field. In
other words, it means the development of multicanonical
statistical mechanics for disordered $1D$ spin chain system in the
external field. In order to do this the classical spin-glass
Heisenberg type Hamiltonian is investigated (\ref{38}). In the
first step, the nonperturbed Hamiltonian (\ref{39}) is used and
the system of algebraic equations (\ref{40}) is found for
computation of all possible stable spin configurations on the spin
chain scale. After this, the statistic weight
$P_\nu({\bf{E}}_\nu;d_0(T))$ for certain energies ${\bf{E}}_\nu$
of stable non-perturbed spin chain is found simply by formula
(\ref{41}). With the help of formulas (\ref{41})-(\ref{46}) the
partition function ${\bf{Z}}({\bf{E}};{\bf{g}})$, which  gives an
energy distribution in the spin chain system after inserting of
external field, is defined (\ref{48}).  The Helmholtz free energy
(\ref{51}) is constructed and it is shown that the $1D$ random
spin chain system subjected to weak external field $E_i$ and the
order $E_c^\pm\varpropto{1/N_x}$ can be severely changed (see Fig.
6).

The mean value of complex interaction   potential
$\langle{\delta{U_{L_{x}}}(x,{\bf{g}})}\rangle_{(\updownarrow)}$
between spin chain and external field is found from formula
(\ref{07}) taking into account expressions (\ref{41}) and
(\ref{48}). An averaging potential is used for computation of
polarizability coefficient related with the orientation effects
(\ref{09}) and correspondingly for generalization of
Clausius-Mossotti equation (\ref{11}). Remind that  Eq. (\ref{11})
has a content only on the  micro-scale space and the nano-scale
time. Note, that the  analysis of Eq. (\ref{11}) together with
(\ref{23}) show that in the spin-glass medium with the help of
weak external field it is possible to form regions (layers) with
different stationary dielectric constants
$\epsilon_{st}({\bf{g}})$ (see Fig 8). As it is easy to understand
the value of stationary dielectric constant strongly depends on
initial electrostatic dielectric constant $\epsilon_s$ of media.
In particular, it is shown that in the spin-glass mediums with
static dielectric constants $\epsilon_s<4$ under the external
field the value of the stationary dielectric constant can be
changed maximum 3-4 times. In the case  $\epsilon_s>4$ analysis
shows that the value of $Re(\Lambda(\textbf{g}))\rightarrow 1$ and
correspondingly stationary dielectric constant can be complex
function $\epsilon_{st}({\bf{g}})$ depending on the parameters of
external field ${\bf{g}}$. Moreover for a set of some parameters
of weak external field it is safe to say that the $ \epsilon_{st}
({\bf{g}})$ can have a large imaginary part (see (\ref{52})).

 Obviously, near to this value critical effecters
are hold. In this case taking into account the relaxation
processes, which are going in the lattice, the system becomes
ordered and is characterized by macroscopic classical
polarization.

The second main step of investigation is the frequency-dependant
dielectric constant or dielectric permittivity. Generalized
Clausius-Mossotti Eq. (\ref{11}) may be used for derivation of
generalized equation for dielectric permittivity (\ref{14})  for
the frequencies range $\omega\gg\Omega$. It is important to say
that the analysis of  stationary dielectric constant value
$\epsilon_{st}({\bf{g}})$  shows that dielectric permittivity
function $\epsilon(w,{\bf{g}})$ (see (\ref{54}) and (\ref{55})\,)
can have  positive, as well as negative dispersions. The last
circumstance is very important especially for the $x$-range
frequency $\omega$ because on the scale of wavelength of external
field  $\lambda_s$ (see Fig. 7) the dielectric permittivity in the
region of positive dispersion "I" and "III"
$Re\,\epsilon_1(\omega,\textbf{g})<1$ whereas in the neighbor
region of negative dispersion  "II" and "IV" correspondingly
$Re\,\epsilon_2(\omega,\textbf{g})>1$ (see for example
\cite{Smith,Pendry,Veselago}). Recall that this occurs when
initial electrostatic dielectric constant $\epsilon_{s}>4$.
Evidently that the difference between dielectric permittivity of
neighboring layers (\ref{57}) in the $x$-ray region as compared
with usual cases when the external field is absent can differ
drastically.

In this case collective orientation effects of spins make main
investment in the permittivity function   (macroscopic
polarization) which radically changes behavior of media on the
space-time scale of external standing electromagnetic wave.
Particularly only a presence of large imaginary part in the
coefficient of $\delta\epsilon_{st} ({\bf{g}})$ brings to an
essential growth of the real part of difference
$\bigl\{i({\omega_p}/{\omega}) +({\omega_p}/{\omega})^2\bigr\}$.

 Thus, obviously  it is possible to make and control the
 periodic structure from dielectric permittivities in the
 spin-glass type dielectric  medium  even with the help of
 weak external standing (\ref{02}) electromagnetic field. In other
 words  a new type of controlled radiator for generation of
 coherent $x$-radiation is suggested.

On the other hand the periodic dielectric permittivity function
may be illustrated as a $1D$ lattice of quantum dots (quantum
strings) which can be interesting in the view of quantum computer
making.

Finally it is important to note that the first time  a new
mathematical method for reducing the problem of dynamical $3D$
disordered spin system in the external field to the two
conditionally separated $1D$ problems is developed. The last
circumstance can be used for elaboration of new extremely
effective parallel algorithms which is very important for
systematic investigations of mentioned problem in the framework of
numerical simulations method.

\section{Acknowledgments}
This work  was partially supported by   Armenian and Taiwan
Science Research Councils. AG also thanks ISTC grant N-655 for
support.

AG  expresses  thanks to Prof. A. R. Mkrtchyan who has told him
about the experiment \cite{Al} and required to investigate the
cause of this phenomenon.

\end{document}